\documentstyle[11pt,a4paper]{amsart}

\newcommand{\rpb}{{R.~P.~Brent}}
\newcommand{\MC}{{\em Mathematics of Computation}}
\newcommand{\CACM}{{\em Comm.\ ACM}}
\newcommand{\lbrk}{{\linebreak[0]}}	%

\theoremstyle{definition}

\begin{document}
\thanks{Copyright \copyright\ 1997, R.~P.~Brent%
	\hfill rpb170tr typeset using {\AmS}-\LaTeX}

\title[A fast normal random number generator]		%
      {A fast vectorised implementation of Wallace's
	 normal random number generator}		%

\author[R.\ P.\ Brent]{Richard P.\ Brent}
\address{Computer Sciences Laboratory\\ %
	RSISE\\ %
	Australian National University\\
	Canberra, ACT~0200\\
	Australia}
\date{14 April 1997}
\email{Richard.Brent@@anu.edu.au}

\thispagestyle{empty}

\begin{abstract}
Wallace has proposed a new class of pseudo-random generators
for normal variates. These generators do not require a stream of uniform
pseudo-random numbers, except for initialisation. The inner loops are
essentially matrix-vector multiplications and are very suitable for
implementation on vector processors or vector/parallel processors
such as the Fujitsu VPP300.
In this report we outline Wallace's idea, consider some variations on it,
and describe a vectorised implementation {\tt RANN4}
which is more than three times faster than its best competitors
(the Polar and Box-Muller methods) on the Fujitsu VP2200 and VPP300.
\end{abstract}

\maketitle

\section{Introduction}
\label{Sec:Intro}

Several recent papers~\cite{And90,rpb132,Pet92,Pet93}
have considered the generation of uniformly distributed
pseudo-random numbers on vector and parallel computers.
In many applications, random numbers from specified non-uniform distributions
are required.
A common requirement is for the normal distribution,
which is what we consider here.
In principle it is sufficient to consider methods
for generating normally distributed
numbers with mean $0$ and variance $1$, %
since translation and scaling can easily be performed to give
numbers with mean $\mu$ and variance $\sigma^2$ (usually referred to as
numbers with the $N(\mu,\sigma^2)$ distribution).

The most efficient methods for generating %
normally distributed random numbers on sequential
machines~\cite{Ahr72,rpb023,Dev86,Gri85,Kin77,Knu81,Lev92,Wal76}
involve the use of different approximations
on different intervals, and/or the use of ``rejection'' methods,
so they do not vectorise well.
Simple, ``old-fashioned'' methods
may be preferable on vector processors.
In~\cite{rpb141} we described two such
methods, the Box-Muller~\cite{Mul59} and Polar methods~\cite{Knu81}.
The Polar method was implemented as {\tt RANN3} and was
the fastest vectorised method for normally distributed numbers known at the
time~\cite{Pet88,Pet93},
although much slower than the best uniform random number
generators. For example, on the Fujitsu VP2200/10
a normal random number using {\tt RANN3}
requires an average of 21.9 cycles,
but a good generalised Fibonacci uniform
random number generator requires only 2.21 cycles.
(A cycle on the VP2200/10 is 3.2 nsec.
Since four floating-point operations %
can be performed per cycle, the
theoretical peak performance of the VP2200/10 is 1250 Mflop.
The cycle time of the VPP300 is 7~nsec %
but the pipelines are wider, so the theoretical peak performance
is 2285 Mflop.)

Recently Wallace~\cite{Wal94} proposed
a new class of pseudo-random generators
for normal variates. These generators do not require a stream of uniform
pseudo-random numbers (except for initialisation)
or the evaluation of elementary functions such as
log, sqrt, sin or cos (needed by the Box-Muller and Polar methods).
The crucial observation is that,
if $x$ is an $n$-vector of normally distributed random numbers,
and $A$ is an $n \times n$ orthogonal matrix, then
$y = Ax$ is another $n$-vector of normally distributed numbers.
Thus, given a pool of $nN$ %
normally distributed numbers,
we can generate another pool of $nN$ normally distributed numbers
by performing $N$ matrix-vector
multiplications. %
The inner loops are
very suitable for
implementation on vector processors such as the VP2200
or vector/parallel processors
such as the VPP300. The vector lengths are proportional to
$N$, and the number of arithmetic operations per
normally distributed number is proportional to $n$.
Typically we choose $n$ to be small, say $2 \le n \le 4$,
and $N$ to be large.

Wallace implemented variants of his new method on a scalar RISC workstation,
and found that its speed was comparable to that of a fast uniform generator.
The same performance relative to a fast uniform generator
is achievable on a vector processor, although some
care has to be taken with the implementation (see~\S\ref{Sec:Imp}).

In~\S\ref{Sec:Wallace} we describe Wallace's new methods in more detail.
Some statistical questions are considered
in~\S\S\ref{Sec:Cons}--\ref{Sec:Discard}.
Aspects of implementation on a vector processor are
discussed in~\S\ref{Sec:Imp},
and details of an implementation on the VP2200 and VPP300 are given
in~\S\ref{Sec:RANN4_imp}.
Some conclusions are drawn in~\S\ref{Sec:Conc}.

\section{Wallace's Normal Generators}
\label{Sec:Wallace}

The idea of Wallace's new generators is to keep a pool of $nN$ normally
distributed pseudo-random variates. As numbers in the pool are used,
new normally distributed variates are generated by forming appropriate
combinations of the numbers which have been used. On a vector processor
$N$ can be large and the whole pool can be regenerated with
only a small number of vector operations\footnote{The process of regenerating
the pool will be called a ``pass''.}.

As just outlined, the idea is the same as that of the generalised Fibonacci
generators for uniformly distributed numbers~-- a pool of random numbers is
transformed in an appropriate way to generate a new pool.
As Wallace~\cite{Wal94} observes, we can regard the uniform,
normal and exponential
distributions as maximum-entropy distributions subject to the constraints:

\begin{itemize}
\item[] $0 \le x \le 1$ (uniform)
\item[] $E(x^2) = 1$ (normal)
\item[] $E(x) = 1$, $x \ge 0$ (exponential).
\end{itemize}

We want to combine $n \ge 2$ numbers in the pool so as to satisfy the
relevant constraint, but to conserve no other statistically relevant
information. To simplify notation, suppose that $n = 2$ (there is no problem
in generalising to $n > 2$). %
Given two
numbers $x$, $y$ in the pool, we could satisfy the ``uniform'' constraint
by forming
\[x' \leftarrow (x + y) \bmod 1,\]
and this gives the family of generalised Fibonacci generators~\cite{rpb141}.

We could satisfy the ``normal'' constraint by forming
\[
\left(\begin{array}{c}
x' \\ y' \end{array}\right) \leftarrow
A \left(\begin{array}{c}
x \\ y \end{array}\right),
\]
where $A$ is an orthogonal matrix, for example
\[
A = {1 \over \sqrt{2}}\left(\begin{array}{cc}
1  &  1 \\
-1 &  1 \end{array}\right)
\]
or
\[
A = {1 \over 5}\left(\begin{array}{cc}
4  &  3 \\
-3 &  4 \end{array}\right).
\]
Note that this generates two new pseudo-random normal variates $x'$ and $y'$
from $x$ and $y$, and the constraint
\[ {x'}^2 + {y'}^2 = x^2 + y^2 \]
is satisfied because $A$ is orthogonal.

Suppose the pool of previously generated pseudo-random numbers contains
$x_0, \ldots, x_{N-1}$ and $y_0, \ldots, y_{N-1}$.
Let $\alpha, \ldots, \delta$ be integer constants.
These constants might be fixed throughout, or they might be varied
(using a subsidiary uniform random number generator) each time
the pool is regenerated.

One variant of Wallace's method generates $2N$ new pseudo-random
numbers $x'_0, \ldots, x'_{N-1}$ and $y'_0, \ldots, y'_{N-1}$ using
the recurrence

\begin{equation}
\left(\begin{array}{c}
x'_j\\			%
y'_j \end{array}\right)
= A
\left(\begin{array}{c}
x_{\alpha j + \gamma \;\bmod\; N}\\
y_{\beta j + \delta \;\bmod\; N}\end{array}\right)	\label{eq:Wrec}
\end{equation}
for $j = 0, 1, \ldots, N-1$.
The vectors $x'$ and $y'$ can then overwrite $x$ and $y$, and be used as the
next pool of $2N$ pseudo-random numbers. To avoid the copying overhead,
a double-buffering scheme can be used.

\section{Desirable Constraints}
\label{Sec:Cons}

In order that all numbers in the old pool $(x,y)$ are used to generate the
new pool $(x',y')$, it is essential that the indices
\[ \alpha j + \gamma \bmod N \]
and
\[ \beta j + \delta \bmod N \]
give permutations of $\{0, 1, \ldots, N-1 \}$
as $j$ runs through $\{0, 1, \ldots, N-1 \}$.
A necessary and sufficient condition for this is that
\begin{equation}
{\rm GCD}(\alpha, N) = {\rm GCD}(\beta, N) = 1.     \label{eq:gcd}
\end{equation}
For example, if $N$ is a power of $2$, then
any odd $\alpha$ and $\beta$ may be chosen.

The orthogonal matrix $A$ must be chosen so each of its rows has at least
two nonzero elements, to avoid repetition of the same pseudo-random numbers.
Also, these nonzeros should not be too small.

For implementation on a vector processor it would be efficient to
take $\alpha = \beta = 1$ so vector operations have unit strides.
However, statistical considerations indicate that unit strides should be
avoided. To see why,
suppose $\alpha = 1$.  Thus, from~(\ref{eq:Wrec}),
\[ x'_j = a_{0,0}x_{j + \gamma \;\bmod\; N} +
	  a_{0,1}y_{\beta j + \delta \;\bmod\; N},\]
where $|a_{0,0}|$ is not very small.
The sequence $(z_j)$ of random numbers returned to the user is
\[
\begin{array}{ccc}
x_0, \ldots, x_{N-1},   &y_0, \ldots, y_{N-1},   &\\
x'_0, \ldots, x'_{N-1}, &y'_0, \ldots, y'_{N-1}, &\ldots
\end{array}
\]
so we see that $z_n$ is
strongly correlated with $z_{n+\lambda}$
for $\lambda = 2N - \gamma$.

Wallace~\cite{Wal94} suggests a ``vector'' scheme where
$\alpha = \beta = 1$
but $\gamma$ and $\delta$ vary at each pass.
This is certainly an improvement over keeping $\gamma$ and $\delta$ fixed.
However, there will still be correlations over segments of length $O(N)$ in
the output, and these correlations can be detected by suitable statistical
tests. Thus, we do not recommend the scheme for a library routine, although
it would be satisfactory in many applications.

We recommend that $\alpha$ and $\beta$
should be different, greater than $1$,
and that $\gamma$ and $\delta$ should be selected
randomly at each pass to reduce any residual correlations.

For similar reasons, it is desirable
to use a different orthogonal matrix $A$
at each pass.
Wallace suggests randomly selecting from two predefined $4 \times 4$ matrices,
but there is no reason to limit the choice to two\footnote{Caution:
if a finite set of predefined matrices is used,
the matrices should be multiplicatively independent over $GL(n,R)$.
(If $n = 2$, this means that the rotation angles should be independent
over the the integers.)
In particular, no matrix should
be the inverse of any other matrix in the set.
}.
We prefer to choose ``random'' $2 \times 2$ orthogonal matrices with
rotation angles not too close to a multiple of $\pi/2$.

\section{The Sum of Squares}
\label{Sec:Stats}

As Wallace points out, an obvious defect of the schemes described
in~\S\S\ref{Sec:Wallace}--\ref{Sec:Cons} is that the sum of squares of the
numbers in the pool is fixed (apart from the effect of rounding errors).
For truly random normal variates
the sum of squares should have the chi-squared distribution
$\chi^2_\nu$, where $\nu = nN$ is the pool size.

To overcome this defect,
Wallace suggests that one pseudo-random number from each pool
should not be returned to the user, but should be used to
approximate a random sample $S$ from the $\chi^2_\nu$ distribution.
A scaling factor can be introduced to ensure
that the sum of squares of the $\nu$ values in the pool (of which $\nu-1$ are
returned to the user)
is $S$. This only involves scaling the matrix $A$, so the inner loops are
not significantly changed.

There are several good approximations to the $\chi^2_\nu$ distribution
for large $\nu$. For example,
\begin{equation}
2\chi^2_\nu \simeq \left(x + \sqrt{2\nu - 1}\right)^2\;, \label{eq:chiapp}
\end{equation}
where $x$ is $N(0,1)$.
More accurate approximations are known~\cite{AS}, but~(\ref{eq:chiapp})
should be adequate if $\nu$ is large.

\section{Restarting}
\label{Sec:Restart}

Unlike the case of generalised Fibonacci uniform random number
generators~\cite{rpb133},
there is no well-developed theory to tell us what the period of the
output sequence of pseudo-random normal numbers is.
Since the size of the state-space is at least
$2^{2wN}$, where $w$ is the number of bits in a floating-point fraction
and $2N$ is the pool size (assuming the worst case $n = 2$),
we would expect the period to be at least
of order $2^{wN}$ (see Knuth~\cite{Knu81}), but it is difficult to
guarantee this. One solution is to restart
after say 1000N numbers have been generated, using
a good uniform random number generator with guaranteed long period
combined with the Box-Muller method to refill the pool.

\section{Discarding Some Numbers}
\label{Sec:Discard}

Because each pool of pseudo-random numbers is, strictly speaking,
determined by the previous pool, it is desirable not to return all
the generated numbers to the user\footnote{Similar remarks apply to some
uniform pseudo-random number generators~\cite{Knu97,Lus94}.}.
If $f \ge 1$ is a constant parameter\footnote{We shall  call $f$ the
``throw-away'' factor.},
we can return a fraction $1/f$ of the generated numbers to the user
and ``discard'' the remaining fraction $(1 - 1/f)$. The discarded numbers
are retained internally and used to generate the next pool.
There is a tradeoff between independence
of the numbers generated and the time required to generate each number which
is returned to the user.  Our tests (described in~\S\ref{Sec:RANN4_imp})
indicate that $f \ge 3$ is satisfactory.

\section{Vectorised Implementation}
\label{Sec:Imp}

If the recurrence~(\ref{eq:Wrec}) is implemented in the obvious way,
the inner loop will involve index computations modulo $N$. It is possible
to avoid these computations. Thus $2N$ pseudo-random numbers can be
generated by $\alpha + \beta - 1$ iterations of a loop of the form
\begin{quotation}
\begin{verbatim}
 	DO J = LOW, HIGH
 	XP(J) = A00*X(ALPHA*J + JX) + A01*Y(BETA*J + JY)
 	YP(J) = A10*X(ALPHA*J + JX) + A11*Y(BETA*J + JY)
 	ENDDO
\end{verbatim}
\end{quotation}
where {\tt ALPHA} $= \alpha$, {\tt BETA} $= \beta$,
and {\tt LOW, HIGH, JX,} and {\tt JY} are integers which are
constant within the loop but
vary between iterations of the loop. Thus, the loop vectorises.
To generate each pseudo-random number requires one load (non-unit
stride), one floating-point add, two floating-point multiplies,
one store, and of order
\[ \alpha + \beta \over N \] %
startup costs.
The average cost should is only a few machine cycles per random
number if $N$ is large and $\alpha + \beta$ is small.

On a vector processor with interleaved memory banks, it is desirable
for the strides $\alpha$ and $\beta$ to be odd so that the maximum
possible memory bandwidth can be achieved. For statistical reasons
we want $\alpha$ and $\beta$ to be distinct and greater than
$1$ (see~\S\ref{Sec:Cons}).
For example, we could choose
\[ \alpha = 3,\;\; \beta = 5, \]
provided ${\rm GCD}(\alpha\beta, N) = 1$ (true if $N$ is a power of~2).
Since $\alpha + \beta - 1 = 7$, the average vector length in
vector operations is about $N/7$.

Counting operations in the inner loop above, we see that generation
of each pseudo-random $N(0,1)$ number requires about two floating-point
multiplications and one floating-point addition, plus one (non-unit stride)
load and one (unit-stride) store.  To transform the $N(0,1)$ numbers to
$N(\mu,\sigma^2)$ numbers with given mean and variance requires an
additional multiply and add (plus a unit-stride load and store)
\footnote{Obviously some optimisations are possible if it is known
that $\mu = 0$ and $\sigma = 1$.}.
Thus, if $f$ is the throw-away factor (see~\S\ref{Sec:Discard}),
each pseudo-random $N(\mu,\sigma^2)$ number returned to the user
requires about $2f+1$ multiplies and $f+1$ additions,
plus $f+1$ loads and $f+1$ stores.

If performance is limited by the multiply pipelines, it might be
desirable to reduce the number of multiplications in the inner loop by
using fast Givens transformations (i.e. diagonal scaling).
The scaling could be undone when the results were copied to the caller's
buffer. To avoid problems of over/underflow, explicit scaling could
be performed occasionally (e.g. once every 50-th pass through the pool
should be sufficient).

The implementation described in~\S\ref{Sec:RANN4_imp}
does not include fast Givens transformations or any particular
optimisations for the case $\mu = 0$, $\sigma = 1$.

\section{RANN4}
\label{Sec:RANN4_imp}

We have implemented the method described
in~\S\S\ref{Sec:Discard}--\ref{Sec:Imp} in Fortran on the VP2200 and VPP300.
The current implementation is called {\tt RANN4}.
The implementation uses {\tt RANU4}~\cite{rpb132}
(or equivalently the SSL2/VPP
{\tt DP\_VRANU4}) to generate uniform pseudo-random numbers
for initialization and generation of the parameters
$\alpha, \ldots, \delta$ (see~(\ref{eq:Wrec}))
and pseudo-random orthogonal matrices (see below).
Some desirable properties of
the uniform random number generator are inherited by {\tt RANN4}.
For example, {\tt DP\_VRANU4} appends the processor number to the seed,
so it is certain that different pseudo-random sequences will be generated
on different processors, even if the user calls the generator with the
same seed on several processors of the VPP300.

The user provides {\tt RANN4} with a work area which must be preserved
between calls.
{\tt RANN4} chooses a pool size of $2N$, where $N \ge 256$
is the largest power of~$2$
possible so that the pool fits within part (about half) of the work area.
The remainder of the work area is used for the uniform generator
and to preserve essential information between calls.
{\tt RANN4} returns an array of normally distributed pseudo-random numbers
on each call.  The size of this array, and the mean and variance of the
normal distribution, can vary from call to call.

The parameters $\alpha, \ldots, \delta$ (see~(\ref{eq:Wrec})) are chosen
in a pseudo-random manner, once for each pool,
with $\alpha \in \{3, 5\}$ and $\beta \in \{7, 11\}$.
The parameters $\gamma$ and $\delta$ are chosen uniformly from
$\{0, 1, \ldots, N-1\}$.
The orthogonal matrix $A$ is chosen in a pseudo-random manner as
\[
A = \left(\begin{array}{cc}
\cos\theta  &  \sin\theta \\
-\sin\theta &  \cos\theta \end{array}\right),
\]
where $\pi/6 \le |\theta| \le \pi/3$ or $2\pi/3 \le \theta \le 5\pi/6$.
The constraints on $\theta$ ensure that
$\min(|\sin\theta|, |\cos\theta|) \ge 1/2$.
We do not need to compute trigonometric functions:
a uniform generator is used to select $t = \tan(\theta/2)$
in the appropriate range, and then $\sin\theta$ and $\cos\theta$
are obtained using a few arithmetic operations.
The matrix $A$ is fixed in each inner loop (though not in each complete pass)
so multiplications by $\cos\theta$ and $\sin\theta$ are fast.

For safety we adopt the conservative choice of throw-away factor
$f = 3$ (see~\S\ref{Sec:Discard}),
although in most applications the choice $f = 2$ (or even $f = 1$) is
satisfactory and significantly faster.

Because of our use of {\tt RANU4} to generate the parameters
$\alpha, \ldots, \delta$ etc,
it is most unlikely that the period of the sequence returned by {\tt RANN4}
will be shorter than the period of the uniformly distributed sequence
generated by {\tt RANU4}.
Thus, it was not considered necessary to
restart the generator as described in~\S\ref{Sec:Restart}.
However, our implementation monitors the sum of squares and
corrects for any ``drift'' caused by accumulation of
rounding errors\footnote{This provides a useful check, because any large
change in the sum of squares must be caused by an error~-- most likely
the user has overwritten the work area.}.

On the VP2200/10, the time per normally distributed number is
approximately $(6.8f + 3.2)$ nsec,
i.e. $(1.8f + 1.0)$ cycles.
With our choice of $f = 3$ this is 23.6 nsec or 6.4 cycles.
The fastest version, with $f = 1$, takes 10 nsec or 2.8 cycles.
For comparison, the fastest method of those considered in~\cite{rpb141}
(the Polar method) takes 21.9 cycles. %
Thus, we have obtained a speedup by a factor of about 3.2 in the case $f = 3$.

Times on the VPP300 are typically faster by a factor
of about two.
For example, the time per normally distributed number is 11.4 nsec if $f = 3$
and 5.4 nsec if $f = 1$.

Various statistical tests were performed on {\tt RANN4} with several
values of the throw-away factor~$f$. For example:

\begin{itemize}
\item If $(x,y)$ is
a pair of pseudo-random numbers with (supposed) normal $N(0,1)$ distributions,
then $u = \exp(-(x^2 + y^2)/2)$ should be uniform in $[0,1]$,
and $v = {\rm{artan}}(x/y)$ should be uniform in $[-\pi/2, +\pi/2]$.
Thus, standard tests for uniform pseudo-random numbers can be applied.
For example, we generated batches of (up to) $10^7$ pairs of numbers,
transformed them to $(u,v)$ pairs, and tested uniformity of $u$
(and similarly for~$v$)
by counting the number of values occurring in $1,000$ equal size bins
and computing the $\chi_{999}^2$ statistic. This test was repeated several
times with different initial seeds etc.
The $\chi^2$ values were not
significantly large or small for any $f \ge 1$.

\item We generated a batch of up to $10^7$ pseudo-random numbers,
computed the sample mean, second and fourth moments,
repeated a number of times,
and compare the observed and expected distributions of sample moments.
The observed moments were not
significantly large or small for any $f \ge 3$.
The fourth moment was sometimes
significantly small (at the 5\% confidence level) for $f = 1$.
\end{itemize}

A possible explanation for the behaviour of the fourth moment when $f = 1$
is as follows. Let the maximum absolute value of numbers in the pool at one
pass be $M$, and at the following pass be $M'$.
By considering the effect of the orthogonal transformations applied to
pairs of numbers in the pool, we see that (assuming $n=2$),
\[ M/\sqrt{2} \le M' \le \sqrt{2}M\;. \]
Thus, there is a correlation in the size of outliers at successive passes.
The correlation for the subset of values returned to the user is reduced
(although not completely eliminated) by choosing $f > 1$.

\section{Conclusion}
\label{Sec:Conc}

We have shown that normal pseudo-random number generators based on
Wallace's ideas vectorise well,
and that their speed on a vector processor is close to
that of the generalised Fibonacci uniform generators, i.e. only a
small number of machine cycles per random number.

Because Wallace's methods are new, there is little knowledge of their
statistical properties. However, a careful implementation
should have satisfactory statistical properties
provided distinct non-unit strides $\alpha$, $\beta$
satisfying~(\ref{eq:gcd}) are used,
the sums of squares are varied as described
in~\S\ref{Sec:Stats}, and the throw-away factor~$f$ is chosen appropriately.
Wallace uses $n \times n$ orthogonal transformations with
$n = 4$, but a satisfactory (and cheaper)
generator is possible with $n = 2$.

The pool size should be fairly large (subject to storage constraints),
both for statistical reasons and to improve performance of the inner loops.

On a vector-parallel machine such as the VPP300, independent streams of
pseudo-random numbers can be generated on each processor, and no
communication between processors is required after the initialization phase.

\subsection*{Acknowledgements}				%
Thanks to Chris Wallace for sending me a preprint of his
paper~\cite{Wal94} and commenting on a preliminary version~\cite{rpb_normalw}
of this report.
Also, thanks to Don Knuth for discussions regarding the properties of
generalised Fibonacci methods and for bringing the reference~\cite{Lus94}
to my attention.
This work was supported in part by a Fujitsu-ANU research agreement.


\begin{thebibliography}{99}

\bibitem{AS}
M.~Abramowitz and I.~A.~Stegun,
{\em Handbook of Mathematical Functions},
Dover, New York, 1965, Ch.~26.

\bibitem{Ahr72}
J.~H.~Ahrens and U.~Dieter,
Computer methods for sampling from the exponential and normal
distributions,
{\em Communications of the ACM} 15 (1972), 873-882.

\bibitem{And90}
S.~L.~Anderson,
Random number generators on vector supercomputers and other advanced
architectures,
{\em SIAM Review} 32 (1990), 221-251.

\bibitem{rpb023} \rpb, Algorithm 488: A Gaussian pseudo-random number
generator (G5),
\CACM\ 17 (1974), 704-706. %

\bibitem{rpb132}
R.~P.~Brent,
Uniform random number generators for supercomputers,
{\em Proc.\ Fifth Australian Supercomputer Conference},
Melbourne, December 1992, 95-104.\\
{\tt ftp://nimbus.anu.edu.au/{\lbrk}pub/Brent/{\lbrk}rpb132.dvi.Z}~.

\bibitem{rpb141}
R.~P.~Brent,
{\em Fast Normal Random Number Generators for Vector Processors},
Technical Report TR-CS-93-04, Computer Sciences Laboratory,
Australian National University, March 1993.
{\tt ftp://nimbus.anu.edu.au/{\lbrk}pub/Brent/{\lbrk}rpb141tr.dvi.Z}~.

\bibitem{rpb133}
R.~P.~Brent, On the periods of generalized Fibonacci recurrences,
\MC\ 63 (1994), 389-401.

\bibitem{rpb_normalw}
R.~P.~Brent,
{\em Wallace's fast normal random number generators: preliminary report},
Fujitsu Area~4 Project Report,
October 1994, 6~pp.

\bibitem{Dev86}
    L.~Devroye,
    {\em Non-Uniform Random Variate Generation.}
    Springer-Verlag, New York, 1986.

\bibitem{Gri85}
P.~Griffiths and I.~D.~Hill (editors),
{\em Applied Statistics Algorithms},
Ellis Horwood, Chichester, 1985.


\bibitem{Kin77}
A.~J.~Kinderman and J.~F.~Monahan,
Computer generation of random variables using the ratio of
uniform deviates,
{\em ACM Transactions on Mathematical Software} 3 (1977), 257-260.

\bibitem{Knu81}
    D.~E.~Knuth,
    {\em The Art of Computer Programming,
    Volume 2: Seminumerical Algorithms} (second edition).
    Addison-Wesley, Menlo Park, 1981.

\bibitem{Knu97}
    D.~E.~Knuth,
    {\em The Art of Computer Programming,
    Volume 2: Seminumerical Algorithms} (third edition).
    Addison-Wesley, Menlo Park, to appear. %

\bibitem{Lev92}
J.~L.~Leva, A fast normal random number generator,
{\em ACM Transactions on Mathematical Software} 18 (1992), 449-453.

\bibitem{Lus94}
M.~L\"uscher,	%
A portable high-quality random number generator for lattice field theory
simulations,
{\em Computer Physics Communications} 79 (1994), 100-110.

\bibitem{Mul59}
    M.~E.~Muller,
    A comparison of methods for generating normal variates on
    digital computers.
    {\em J.~ACM} 6:376--383, 1959.

\bibitem{Pet88}	%
    W.~P.~Petersen,
    Some vectorized random number generators for uniform,
    normal, and Poisson distributions for CRAY X-MP,
    {\em J.~Supercomputing}, 1:327--335, 1988.

\bibitem{Pet92}
    W.~P.~Petersen,
    {Lagged Fibonacci Series Random Number Generators for the
    NEC SX-3},
    {\em International J.\ of High Speed
    Computing} 6 (1994), 387-398.

\bibitem{Pet93}
    W.~P.~Petersen,
    {\em Random Number Generators on Vector Architectures},
    preprint, 1993.

\bibitem{Wal76}
C.~S.~Wallace,
Transformed rejection generators for Gamma and Normal pseudorandom variables,
{\em Australian Computer Journal} 8 (1976), 103--105.

\bibitem{Wal94}
C.~S.~Wallace,
Fast Pseudo-Random Generators for Normal and Exponential Variates,
{\em ACM Trans.\ on Mathematical Software} 22 (1996), 119--127. %

\end{thebibliography}
\end{document}